\documentstyle[epsf,rotate]{mn}
\newcommand{\abs}[1]{|#1|}
\newcommand{\be}{\begin{equation}}
\newcommand{\ee}{\end{equation}}

\def\Msol{{M_\odot}}
\def\kpc{{\rm kpc}}
\def\spose#1{\hbox to 0pt{#1\hss}} 
\def\lta{\mathrel{\spose{\lower 3pt\hbox{$\sim$}}\raise 2.0pt\hbox{$<$}}}
\def\gta{\mathrel{\spose{\lower 3pt\hbox{$\sim$}}\raise 2.0pt\hbox{$>$}}}
\def\sech{{\rm \,sech}}

\begin{document}

\title[Fat Disks]{LMC Microlensing and Very Thick Disks}
\author[Geza Gyuk and Evalyn Gates]{Geza Gyuk$^1$ and Evalyn Gates$^{2,3}$ \\ 
\\
$^1$S.I.S.S.A., via Beirut 2--4, 34014 Trieste, Italy \\
$^2$Adler Planetarium,1300 Lake Shore Drive, Chicago, IL~~60605\\
$^3$Department of Astronomy \& Astrophysics, The University of Chicago, 
Chicago, IL~~60637}
\date{Received ***}
\maketitle

\begin{abstract}
We investigate the implications of a very thick (scale height 1.5 - 3.0
kpc) disk population of MACHOs. Such a population represents a reasonable
alternative to standard halo configurations of a lensing population.  We
find that very thick disk distributions can lower the lens mass estimate
derived from the microlensing data toward the LMC, although an average
lens mass substantially below $0.3\Msol$ is unlikely.  Constraints from
direct searches for such lenses imply very low luminosity objects: thus
thick disks do not solve the microlensing lens problem. We discuss further
microlensing consequences of very thick disk populations, including an
increased probability for parallax events.
\end{abstract}


\section{Introduction}

Current data from the MACHO collaboration \cite{MACHOmass} indicate that
in the context of a spherical isothermal model with a Maxwellian velocity
distribution, some significant fraction of the Galactic halo is composed
of MACHOs with masses roughly in the range 0.1 to 1.0 $\Msol$. Such masses
are consistent with several astrophysical candidates for MACHOs -- white
dwarfs, neutron stars, and black holes -- each of which presents serious
challenges for stellar formation and evolution theories. However, the
MACHO component of the halo, if it is not the major component, as in Cold
Dark Matter scenarios, may have a very different distribution from the
typically assumed spherical isothermal model. The MACHO distribution may
be in a significantly flattened halo and/or, due to dissipation, more
centrally condensed. In addition, such a distribution might have a
significant rotational component. These possibilities have strong
implications not only for the MACHO fraction of the halo, but also for the
mass estimates derived from the event durations.

The velocity dispersion and mass density distribution which describe the
halo model are crucial input parameters in extracting an estimate of the
lens mass from the data. The event duration is given by the radius of the
Einstein ring divided by the MACHO velocity transverse to the line of
sight, where the Einstein ring radius is a function of the position of the
MACHO along the line of sight and the lens mass. The masses of the lenses
can only be determined statistically, in the context of an a priori
assumption about the distribution and velocity dispersion of the
lenses. Such an analysis, using a spherical isothermal model yields a
central mass estimate of $\approx 0.4\Msol$.

However, as discussed above, the MACHO distribution and velocity
dispersion may be very different from that assumed in the standard halo
model.  Earlier work \cite{us_rotate} in exploring models with a highly
flattened halo and a bulk rotational component has suggested that a very
highly condensed model for the MACHO distribution, such as a thick disk,
might reduce the MACHO mass estimate from the current data to a level
consistent with brown dwarf candidates.  Previous explorations
\cite{us_nomacho} have indicated that such models may be able to reproduce
the observed optical depths toward the Large Magellanic Cloud (LMC) and
the Galactic bulge.

The overall shape of the Galactic halo is unknown.  Attempts to determine
the shape of galactic halo potentials from flaring of the outer Galactic
gas layer \cite{Olling,Sackett} point to a flattened halo; flattening
of the potential is also supported by simulations of the cold dark matter
halo formation \cite{nbody}.  While it is highly unlikely that the entire
halo is in a disk-like configuration, it is not unreasonable to assume that
the MACHO component of the halo is significantly more flattened than the
dark matter halo.  The condensation of a gaseous halo component to form a
very thick disk is likely to result in significant star formation and thus
the production of a population of lens candidates.

In this paper we explore in detail the consequences of such a lens
population.  We first define the density and velocity structure, and
outline the constraints on these distributions.  We then present
predictions and observational consequences of such thick disks; in
particular we determine the expected frequency of parallax events.

\section{Very Thick Disks}
\subsection{Density}

We consider a MACHO population which is distributed in a very thick (fat) 
disk, whose scale height ranges from 1.5 to 3.0 kpc.
The radial profile of these fat disks is assumed to have one of two forms: 
\begin{eqnarray}
\Sigma(r) & = & \Sigma_0 \exp((R_0-R)/r_d) \nonumber \\
\Sigma(r) & = & \Sigma_0 \frac{R_0+a}{R+a}. 
\end{eqnarray}
The first, an exponential disk with scale length $r_d$, is similar to the
known thick disk population, though as we shall see the surface densities
must be much larger. The other, a Mestel disk with core $a$, is closer to
a very flattened halo.

Such distributions were considered by Gates et al. (1997) in the context
of comprehensive Galactic models.  The relevant uncertainties in the
parameters that describe the various components of the Galaxy and
observational constraints on the rotation curve were incorporated in
determining viable models.  In this paper we focus primarily on the disk
component of these models in order to study the implications of such a
lens population in detail and only the disk parameters relevant to our
conclusions will be discussed.

Since the scale length of the disk is not well known, we consider both 3.0
kpc and 4.0 kpc \cite{maxdisk}. While the core radius for a possible
Mestel disk is completely unconstrained, our conclusions are not strongly
dependent on the core radius and therefore we choose $a=3.0$ kpc, a value
which does not violate any current rotation curve contraints.

Profiles of the known thin and thick disk populations can be fit
reasonably well with a theoretically motivated (isothermal) sech$^2$
variation in the vertical direction. We thus assume this variation in our
models as well. We have, finally, for our model volume densities
\begin{eqnarray}
\rho(r,z) & = & \frac{\Sigma_0}{2 h_z} \exp((R_0-R)/r_d) \sech^2(z/h_z)
\nonumber \\ \rho(r,z) & = & \frac{\Sigma_0}{2 h_z} \frac{R_0+a}{R+a}
\sech^2(z/h_z).
\end{eqnarray}
We allow both $\Sigma_0$ and $h_z$ to vary in our consideration of models.

\subsection{Velocity Structure}

Observations of local stars show that the velocity distribution of a
homogenous population of stars in the disk can be adequately represented
by an anisotropic gaussian, possibly with a bulk motion with respect to
the local standard of rest. We therefore follow many other investigations
in taking this as the form for our disk distribution function.

For a flat isothermal disk
\be
\sigma_z^2 = 2 \pi G \rho_0 h_z^2 = \pi G \Sigma_0 h_z.
\ee
For the range of $\Sigma_0 (\sim 100\Msol/pc^2)$ and $h_z (\sim 2.0$ kpc)
that we will be considering, this results in $\sigma_z \sim 50$ km/s. If
stars in the known thick disk population are considered tracers (with a
slightly smaller scale height) for the fat disk, we see that this is
consistent with the measured vertical velocity dispersion of 35-60 km/s
\cite{ojha,casertano}. We therefore adopt equation 4 for our calculations of
$\sigma_z$ for the fat disk. In any event, calculations show that small
deviations from this relation do not effect our results.

Measurements of the velocity ellipsoid for many varieties of stars, both
thick and thin disk, show
two fairly constant characteristics (see e.g. \cite{BnT}):
\begin{eqnarray*}
\sigma_r^2 & \approx & 2 \sigma_z^2 \\
\sigma_\phi^2 & \approx & \sigma_z^2.
\end{eqnarray*}
Finally, populations with high velocity dispersions have been shown to
rotate the galaxy more slowly than those with small dispersions. This
asymmetric drift can be roughly quantified (on obserational and
theoretical grounds, see for example \cite{BnT} or \cite{ojha}) as
\be
v_c-\tilde{v}_\phi = \frac{\sigma_r^2}{120 {\rm km/s}}
\ee
where we assume $v_c=220$ km/s. Thus we take as our distribution function
\be
f=\frac{\rho(r,\phi,z)}{m}
\frac{1}{\sqrt{(2\pi)^3}\sigma_r\sigma_\phi\sigma_z}
e^{-\left[\frac{v_r^2}{2\sigma_r^2}+\frac{(v_\phi-\tilde{v}_\phi)^2}{2\sigma_\phi^2}+\frac{v_z^2}{2\sigma_z^2}\right]}
\ee
where $\sigma_r$, $\sigma_\phi$, $\sigma_z$, and $\tilde{v}_\phi$ vary
with position as discussed above.

\section{Constraints}

\subsection{Disk Surface Density}
Studies of the distribution of tracer stars perpendicular to the disk put
limits on the total (luminous + dark) disk surface density of roughly
$90\Msol/{\rm pc}^2$ \cite{BFG92}. However, such studies can only
constrain the column density out to approximately the height of the tracer
population (1.0 kpc). We require $\Sigma_{tot,1.0}<90\Msol/{\rm pc}^2$ for
our fat disk models. The known thin and thick disk populations, M stars,
gas, bright stars, dust etc., give a column density of at most $\approx 50
\Msol/{\rm pc}^2$ \cite{bahcall,GBF97}, essentially all of which is at a
height $\abs{z}<1.0$kpc. We are left then with
$\Sigma_{fat,1.0}<40\Msol/{\rm pc}^2$. For a sech$^2$ disk we have
	   \be
		\Sigma_0(z)=\Sigma_0 \tanh(z/h_z).
	   \ee
The constraint for $z=1.0 \kpc$ is plotted in Figure 1 as a dotted
line. For short scale lengths this is a stringent limit. By $h_z\approx
1\kpc$, however, the constraint on the surface density has loosened
considerably and beyond  $h_z=2\kpc$ very little can be said about the surface
density and the ``disk'' has become more like an exponential halo.

\begin{figure}
\epsfysize=10.0cm
\centerline{
\rotate[r]{\epsfbox{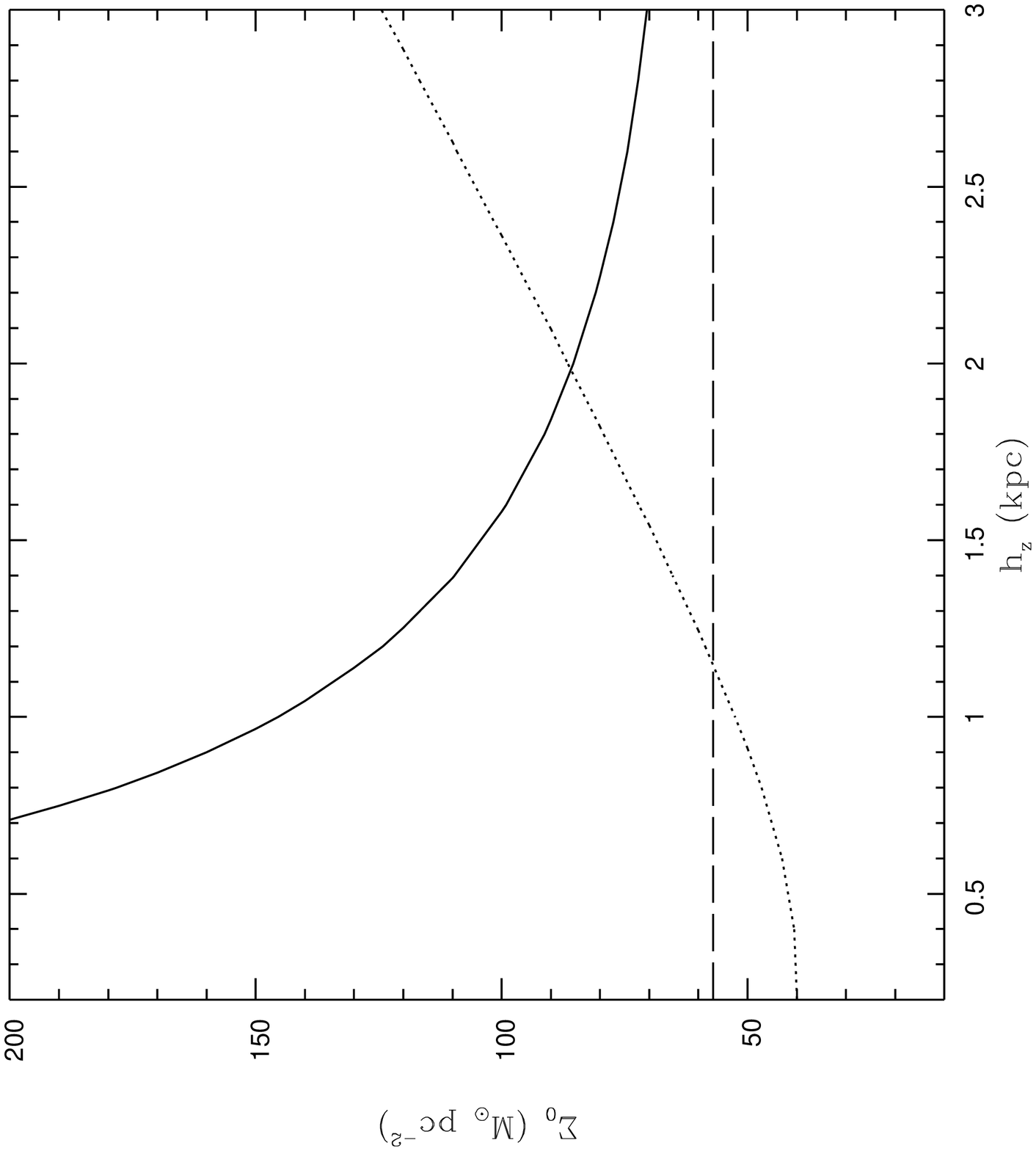}}}
\epsfysize=10.0cm
\centerline{
\rotate[r]{\epsfbox{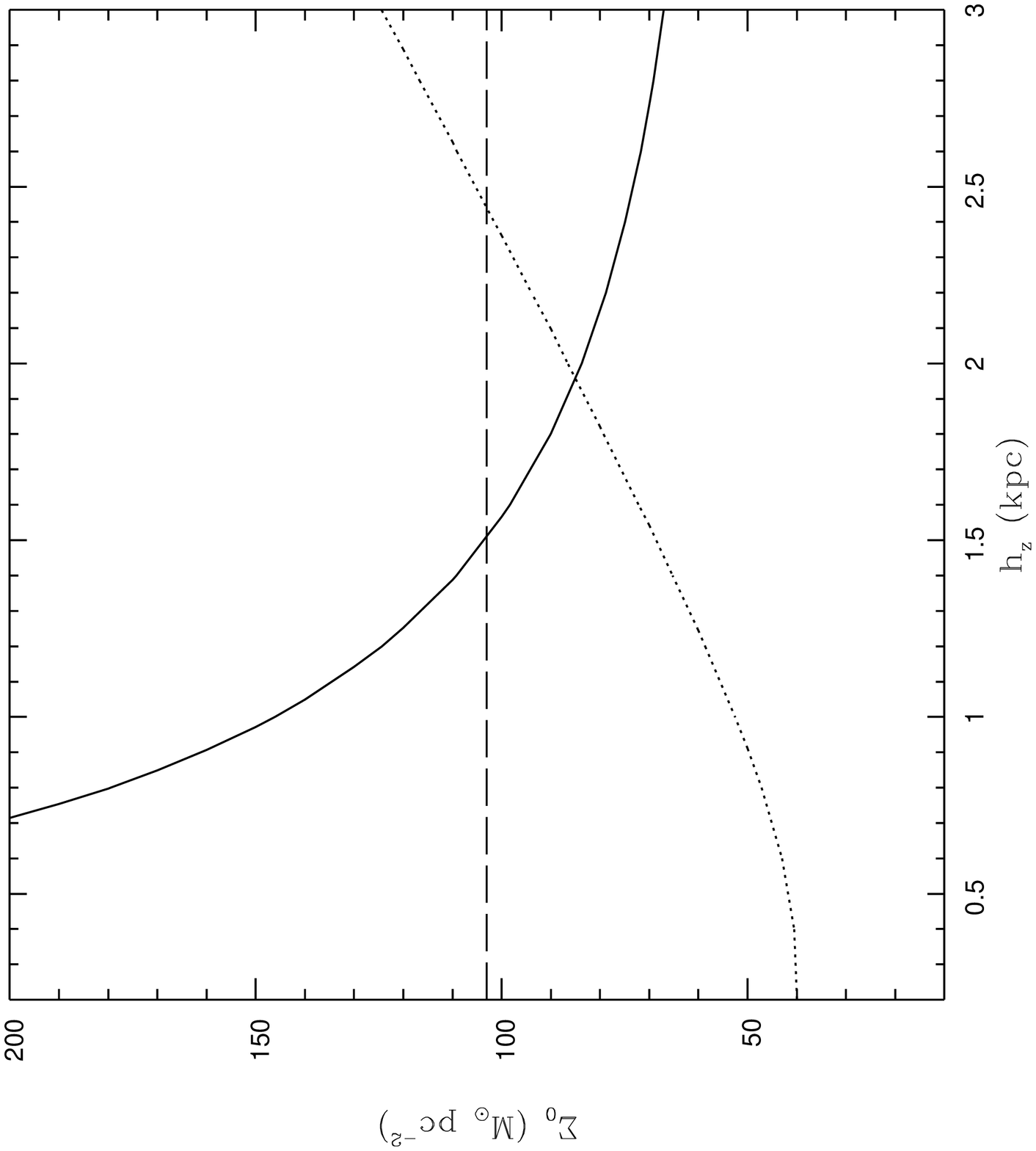}}}
\epsfysize=10.0cm
\centerline{
\rotate[r]{\epsfbox{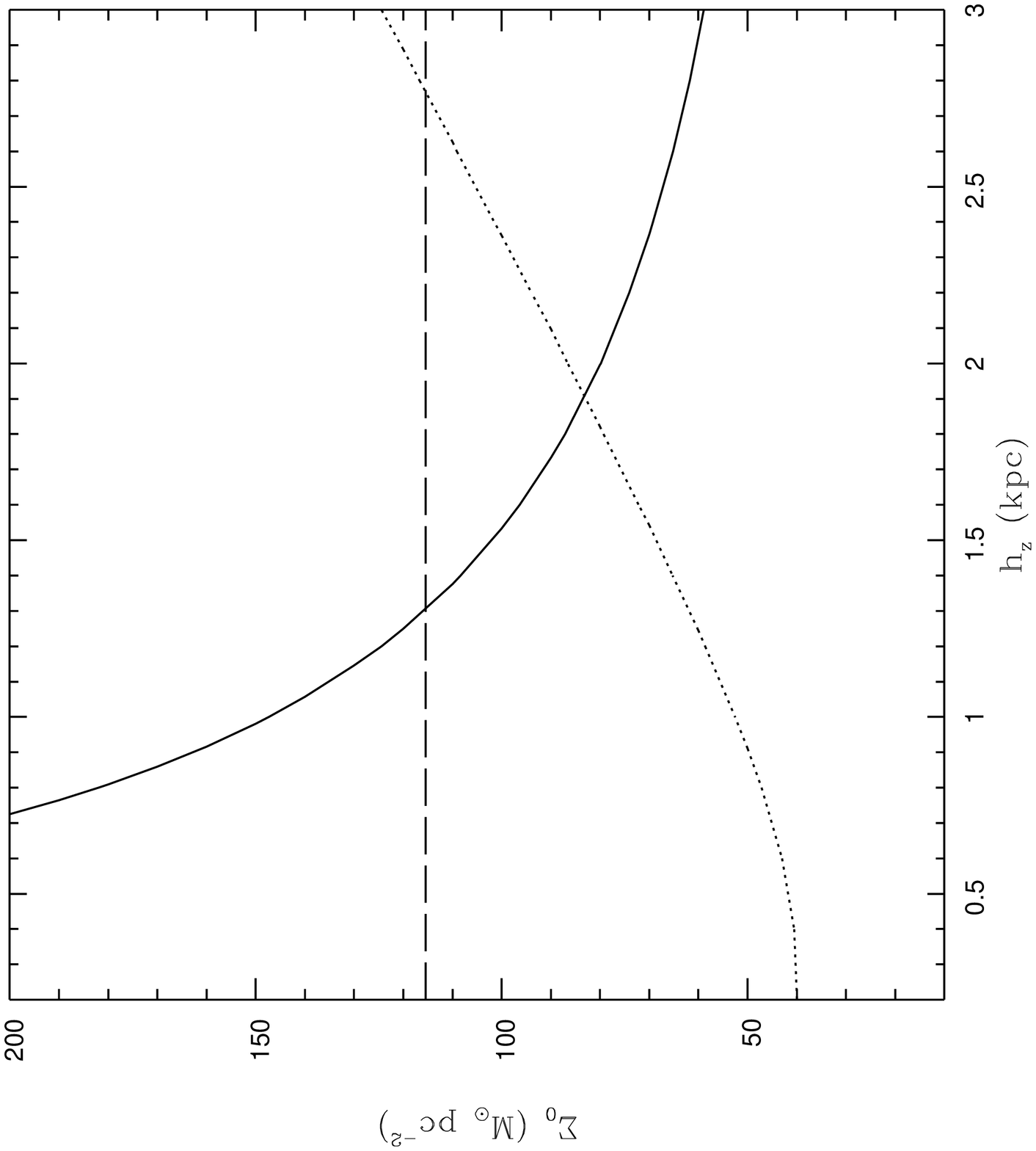}}}

\caption{Constraints on very thick disks:
The region above the dotted line is excluded due to the local 
surface density constraint ($\Sigma_{tot,1.0}<90\Msol/{\rm pc}^2$);
above the dashed line is excluded due to rotation curve constraints;
and below the solid line the LMC optical depth is less than 
$1.0\times 10^{-7}$ for (a) the 3.0 kpc exponential disk, (b) the 4.0 kpc
exponential disk, (c) the Mestel disk.}
\end{figure}

\subsection{Rotation Speed}	
Measurements of the rotation curve and Oort constants are complicated by
our position within the Galaxy. As a very conservative constraint, we adopt the
following. We require the rotation speed of the disk at $R_0$,
$v_c(R_0)<240{\rm km/s}$.  In the calculation of the circular speed we include
known dynamically important components, the bulge and luminous disk,
as well as the fat disk. The bulge mass is taken to be $2 \times 10^{10}\Msol$. 
We assume an exponential radial profile for the luminous thin disk  with, 
as discussed above, $\Sigma_0=50\Msol/{\rm pc}^2$ and
$r_d=3.0$ or 4.0 kpc depending on the model. 

In the thin disk approximation the rotation curve for an exponential disk is
\be
v_E^2(R)=4\pi G \Sigma_0 e^{R_0/r_d} r_d y^2[I_0(y)K_0(y)-I_1(y)K_1(y)]
\ee
where $y=R/2r_d$. For the Mestel disk in the coreless approximation ($a=0$) we have
very simply
\be
v_M^2(R)=2\pi G \Sigma_0 R_0,
\ee
which gives a flat rotation curve. We calculate the bulge component of
the rotation curve as for a Kepler point source. 
Thus we ask that
\be
v_{\rm lum disk}^2(R_0) + v_{\rm bulge}^2(R_0) + v_{\rm MACHOs}^2(R_0) <
(240 {\rm km/s})^2
\ee
where we have left out the halo component to be as generous as possible.
The luminous disk and the bulge together contribute about 160--180km/s leaving
only $\approx 180$km/s maximum for the MACHOs. This translates into
$\Sigma_0< 60, 105, 115 \Msol/pc^2$ for the exponential disk with scale
lengths 3kpc, and 4kpc, and the Mestel disk respectively.  These limits
are plotted as horizontal dashed lines in Figure 1, independent of
$h_z$. Note that for high $h_z$ this limit is much stronger than the
column density limits.

\subsection{Optical Depth}

The optical depth to microlensing is given by 
\be
\tau=\frac{4 \pi G}{c^2 D_s}\int_0^{D_s} x (D_s -x) \rho(x) dx
\ee
where $D_s$ is the distance to the source and $x$ is the observer-lens
distance.  The MACHO collaboration has reported an optical depth toward
the LMC of $2.1^{+1.1}_{-0.7} \times 10^{-7}$ ($2.9^{+1.4}_{-0.9} \times
10^{-7}$) corresponding to 6 (8) microlensing events in their 2 year data
\cite{MACHOmass}.  We consider a lower limit to the optical depth
predicted by our models of $1.0\times 10^{-7}$, approximately one and a
half sigma below the MACHO 6-event data.  This limit is plotted as a solid
line in Figure 1. Because most of the lensing in this class of models
takes place close to the observer where the microlensing tube is narrow,
high surface densities are required. Further, as the scale height is
decreased, the lensing moves closer to the observer where the Einstein
radius is smaller and thus it is very difficult to produce enough lensing
even with extremely high surface densities.

The first 3 constraints are shown in Figure 1.  A few points are
immediately obvious. First, the limit on the column density within $\pm
1.0$kpc is independent of the disk model since it is purely a local
measure. Second, the optical depth is only weakly dependent on the precise
model for the disk with the dependence becoming stronger as the scale
height increases. This again reflects the local nature of the
microlensing: since the fall off above the plane of the disk is
exponential for small scale heights most of the microlensing occurs close
to the observer where the radial dependence of the density is relatively
unimportant. As the scale height increases however, the microlensing
increasingly samples regions distant from the observer where the radial
coordinate is substantially different. The most important difference
between models is the upper limit on the surface density from the rotation
constraints, which varies from $60 \Msol pc^{-2}$ for the 3.0kpc
exponential disk to $115 \Msol pc^{-2}$ for the Mestel disk. Short scale
length exponential disks have much more mass within the solar circle for a
given value of $\Sigma_0$.

As expected, models with scale heights smaller than 1 kpc are thoroughly
ruled out.  Thin or thick disks cannot provide sufficient microlensing
optical depth toward the LMC.  Microlensing is so inefficient for these
small scale heights that column density constraints are as much as an
order of magnitude lower than needed for $\tau_{\rm LMC}$. As the scale
height increases, however, microlensing becomes more efficient and the
required surface density decreases to meet the column density constraints
at about $h_z=2.0 {\rm kpc}$, $\Sigma_0=80 \Msol pc^{-2}$ in all
models. This is where the rotation constraints figure most strongly. For
the more highly condensed model ($r_d=3.0 {\rm kpc}$), the density
increases rapidly towards the center resulting in a higher rotation
velocity for a given surface density. Thus taken together, the three
constraints eliminate the short scale length models entirely and restrict
the distributed models (long scale length exponential and mestel) to a
small allowed region with scale heights $h_z \approx 2-3$kpc and surface
densities, $\Sigma_0 \approx 70-100 \Msol pc^{-2}$.

\subsection{HST and Luminosity Constraints}

In these fat disk models microlensing takes place closer to the observer
than in the standard halo models and thus where the microlensing tube is
narrower. To obtain the same optical depth the density locally must
therefore be greater. It is thus of interest to examine if searches for
faint white dwarfs in the Hubble Deep Field can place significant limits
on such models.  Flynn et al.  (1996) examined the HDF for objects fainter
than $m_I=24.63$ and redder than $V-I=1.8$ down to a limiting magnitude of
$m_I=26.3$. They found no such objects in the $\Omega=3.72\times10^{-7}$
steradian field of the HDF. For an object of I-band absolute magnitude
$M_I$ the volume probed is thus
\be
V=0.9 \frac{\Omega}{3} 10^{[3+0.6(m_I-M_I)]} pc^3.
\ee
Even for relatively bright objects the maximum distance probed is not very
large. Assuming a constant density the number of MACHOs expected in this
volume is then 
\be
N=0.9 \frac{\rho_0}{m} \frac{\Omega}{3} 10^{[3+0.6(m_I-M_I)]} \leq 3
\ee
where the $\leq 3$ is to be consistent with the non-detection of such
objects. For $\rho_0=\Sigma_0/2h_z$ this yields 
\be
M_I> \frac{5}{3} \log_{10}\left[\frac{\Sigma_0}{2h_zm}\right] +18.92.
\ee
with $\Sigma_0$ in $\Msol pc^{-3}$, m in $\Msol$ and $h_z$ in pc.

For a given mass we can then calculate the minimum magnitude for MACHOs to
avoid detection in the HDF as a function of both $\Sigma_0$ and $h_z$.
However, for any single mass, this procedure will not be consistent for
all combinations of $\Sigma_0$ and $h_z$ since this mass may be unlikely
or even ruled out for those values. Instead for each combination we use
the mass estimated from the microlensing event durations (see next section
for these calculations). A contour plot of the minimum magnitudes thus
obtained is shown in Figure 6. Magnitudes for the allowed region are $M_I
\approx 16-17$. Comparison of the local volume density for a typical fat
disk model, $0.02 \Msol pc^{-3}$, to the volume density/age relationship
presented by \cite{Graffetal} shows that the MACHOs must be at least 13
Gyr old and more likely in the 15-17 Gyr range. The fat disk must have
formed in the very earliest stages of the formation of the Galaxy.

\begin{figure}
\epsfysize=10.0cm
\centerline{
\rotate[r]{\epsfbox{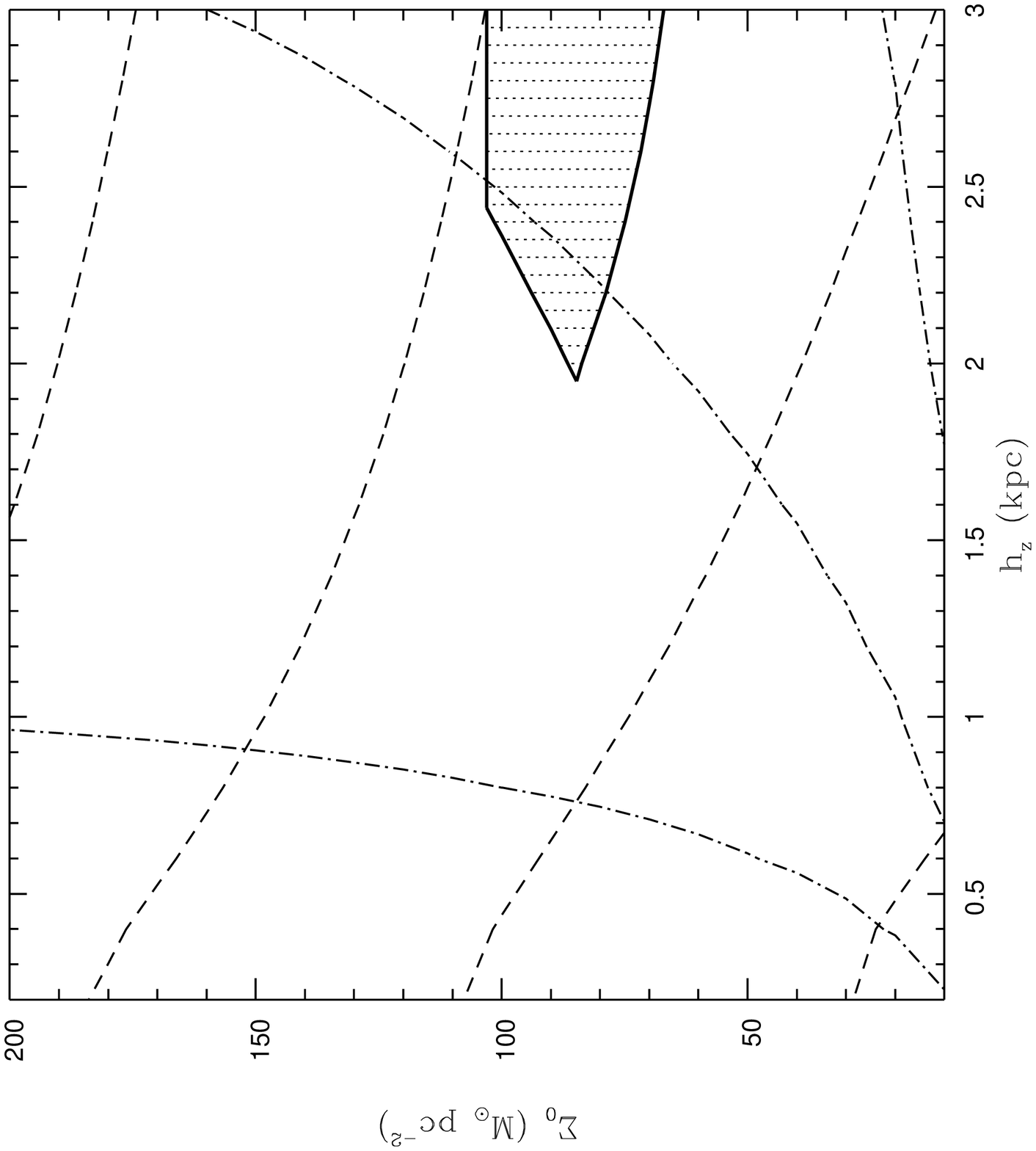}}}
\epsfysize=10.0cm
\centerline{
\rotate[r]{\epsfbox{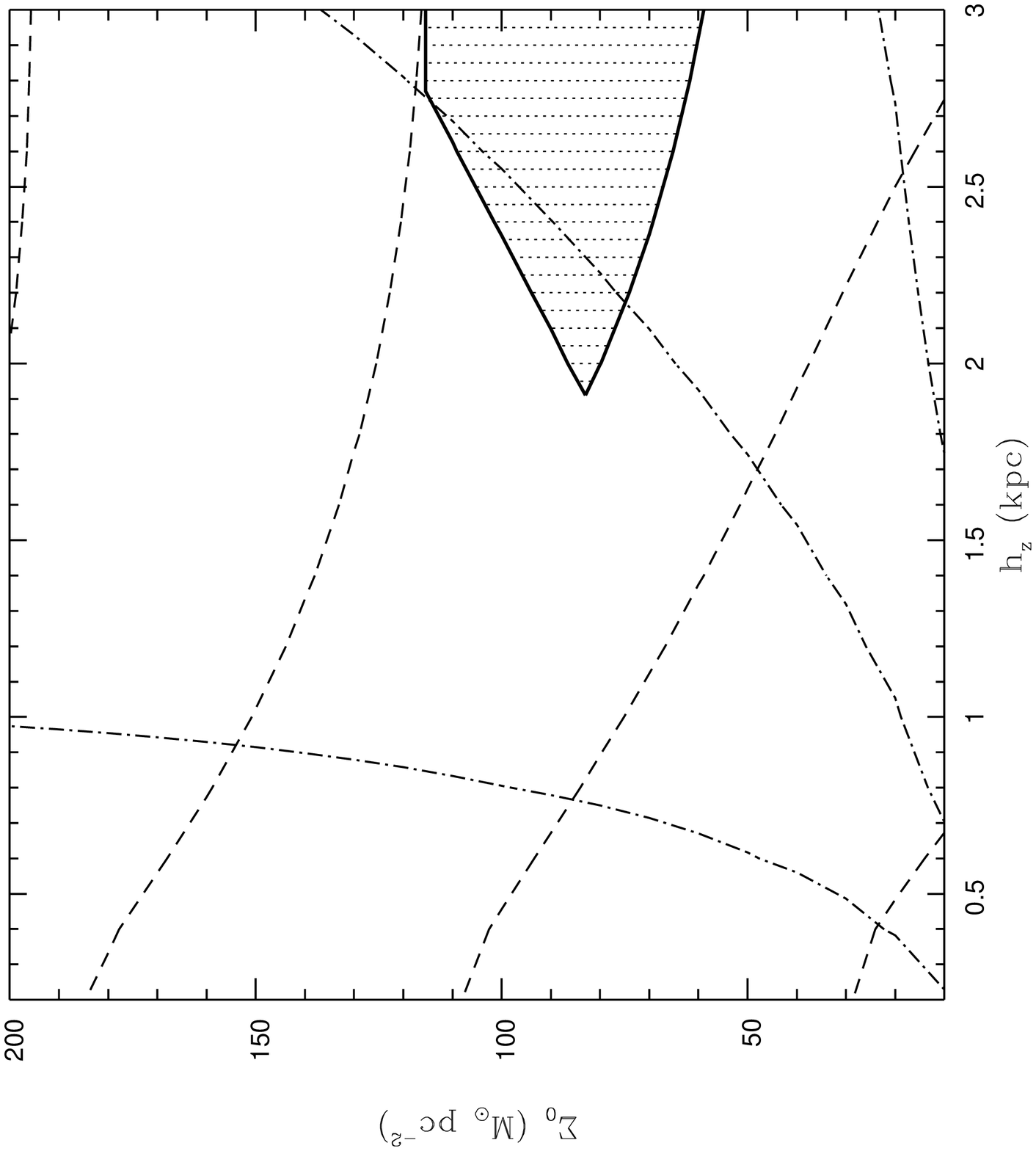}}}
\caption{Mass and luminosity constraint contours:
Contours are I=16,17 and 18 from the right and m=0.1, 0.2, 0.3 and 0.4
$\Msol$ from the bottom. The shaded region corresponds to the range of
$\Sigma_0$ and $h_z$ not excluded by the constraints shown in Figure
1. (a) the 4.0kpc exponential disk. (b) the Mestel disk}
\end{figure}

\section{Predictions and Observational Consequences}
	
\subsection{Microlensing Rate and MACHO masses}

The assumption of the velocity structure of the fat disks we are
considering allows us to calculate not only the optical depth, but also
the microlensing rate to the LMC. In combination with the optical depth we
can find the expected average duration for events for a given model,
\be
\bar{t}_{model}=\frac{\tau}{\Gamma}.
\ee
However, unlike the optical depth, the rate depends on the mass of the
MACHOs, 
\be
\Gamma = \Gamma_{1M_\odot}\sqrt{\frac{1 M_\odot}{\overline{m}}}
\ee
and hence this average duration is also a function of
mass. Comparing the calculated durations to the average observed duration,
$\approx 60$ days we find,
\be
\frac{M_{\rm est}}{1 M_\odot}= \left[\bar{t}_{obs}
\frac{\Gamma_{1M_\odot}}{\tau_{model}}\right]^2.
\ee
where the rate is calculated for $1\Msol$ MACHOs. Contours of this mass
estimate for the range of $\Sigma_0$ and $h_z$ are shown in Figure 2. We
see that the estimated MACHO mass is $\sim 0.3\Msol$ for the entire
allowed region regardless of the type of disk. Fat disks are not a panacea
for the MACHO mass estimate problem.  We calculate the distribution of
Einstein crossing times for our models as a function of MACHO mass and
from this we can calculate the probability for a given MACHO mass to
produce the observed distribution of event crossing times. We show in
Figure 3 the probability as a function of mass for a Mestel disk with
$\Sigma=90\Msol {\rm pc}^-$, $h_z=2.5\kpc$.  Brown dwarf masses, while not
ruled out at the two sigma level, are unlikely. It is generically true for
all of the models we examine that brown dwarf masses are no more likely
for than for the standard halo models. This is so because although the
average predicted mass is smaller, the dispersion is also slightly smaller
and thus the probability for low masses is about the same.

\begin{figure}
\epsfysize=10.0cm
\centerline{
\rotate[r]{\epsfbox{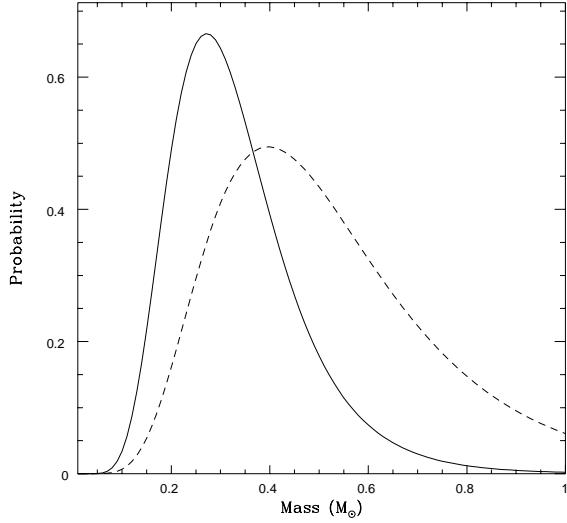}}}
\caption{Probability vs MACHO mass for Mestel disk (solid line) and 
standard halo (dashed line).}
\end{figure}

\subsection{Total Mass in MACHOs}
The analysis of the LMC microlensing events in the context of extended
halos yields a robust estimate of the total mass in MACHOs within 50 kpc
of $M_{\rm baryons} \approx 2 \times 10^{11} \Msol$ in order to produce an
LMC optical depth of about $2 \times 10^{-7}$.  The production of $2
\times 10^{11} \Msol$ in white dwarfs requires a much greater metal
abundance than is seen unless somewhat ad-hoc efficient galactic winds are
invoked to blow the metals out into the intergalactic medium. As discussed
in Gates et al. 1998, MACHOs in a very thick disk configuration can reduce
this mass estimate, but only slightly.  The total mass in a typical very
thick disk is $6.7 \times 10^{10} \Msol$, which results in an optical
depth of about $1.3 \times 10^{-7}$.
This is approximately 1/2 of the mass that would be required for a halo
distribution of MACHOs which would produce the same optical depth.
Basically, this reduction can be understood because most
microlensing is due to lenses within about $20 \kpc$ of the Sun for either
configuration.  The very thick disk has less mass beyond that distance
than a halo.  In addition, we note that since the halo of the Galaxy
extends far beyond the LMC, a standard halo distribution of MACHOs must be
abruptly cut off at $50 \kpc$ in order to avoid a much larger total mass
in MACHOs which would be even more difficult to reconcile with a white
dwarf scenario.

\subsection{$\bf \tau_{\rm SMC}/\tau_{\rm LMC}$}
Strong flattening of the microlensing population also leaves its mark on
the variation of optical depth as a function of direction
\cite{sackettgould,JoshRoman}.  In Figure 4 we show
$\alpha=\tau_{SMC}/\tau_{LMC}$ as a function of $h_z$. For very flat disks
($h_z \approx 0.3$kpc) $\alpha$ is the same for both models. All the
lensing is local, the global configuration of the MACHOs is irrelevant,
and the limiting formula for lensing through a thin disk can be
applied. This predicts a value $\tau_{\rm SMC}/\tau_{\rm LMC}=
{sin^2(b_{\rm LMC})}/{sin^2(b_{\rm SMC})} = 0.60$. As $h_z$ increases
however,the configuration becomes less flat and lensing occurs further
from the observer. Since the line of sight towards the SMC probes closer
to the galactic center, with increasing scale height the SMC direction
passes through denser material than the LMC line of sight and so $\alpha$
increases to about $1.0$ by $h_z=3.0$kpc for the exponential disk
model. For the Mestel disk the density increases less strongly with
decreasing radius and so this effect is not as large, with $\alpha$
reaching only $0.75$ for $h_z=3.0$kpc. By comparison, for a standard
spherical halo $\alpha\approx 1.5$; for a flattened (E6) halo
$\alpha\approx 1.0$.

\begin{figure}
\epsfysize=10.0cm
\centerline{
\rotate[r]{\epsfbox{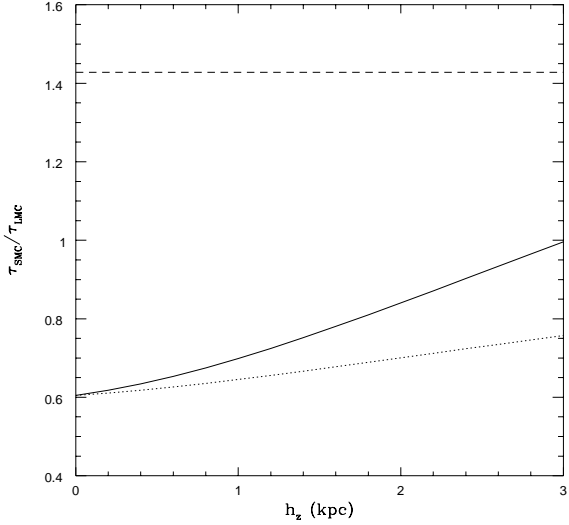}}}
\caption{$\tau_{\rm SMC}/\tau_{\rm LMC}$ for 4.0kpc exponential disk
(dotted line),  Mestel disk (solid line) and standard halo (dashed line).}
\end{figure}

The MACHO and EROS collaborations have recently announced the 
first detection of a microlensing event towards the SMC.  
While it is not possible to draw conclusions from a single event,
an estimate of the ratio of $\tau_{SMC}/\tau_{LMC}$ may be determined 
within the next 5 years or so.

\subsection{Duration and Distance Distribution of Events}
The distribution of event durations for a typical fat disk model is shown
in Figure 5 as a solid line. For comparison, the distribution for a
standard isothermal halo is also shown as a dashed line. It is clear that
distinguishing between these models on the basis of event durations will
require many events. The distribution of events with distance given in
Figure 6 shows more clearly the differences between the two models. Very thick
disk models concentrate the lensing much closer to the observer with a
typical distance of perhaps 5 kpc. In the standard halo case the typical
distance is closer to 15 kpc. Although (as discussed in the next section)
the fractional rate of parallax events is small, it is possible that the
two models could be distinguished with relatively fewer events based on
the distances derived.

\begin{figure}
\epsfysize=10.0cm
\centerline{
\rotate[r]{\epsfbox{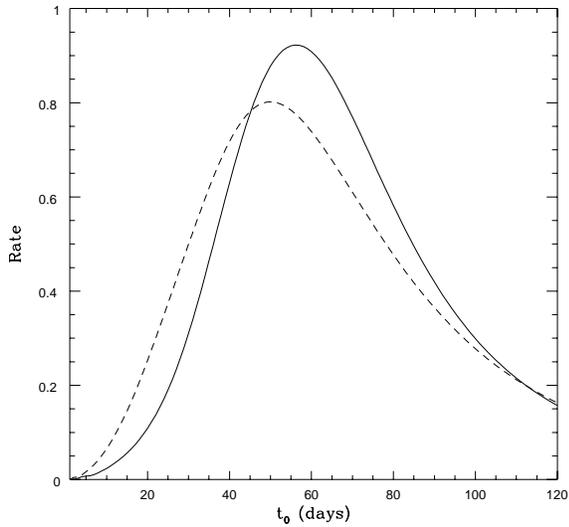}}}
\caption{Distribution of event durations for a typical thick disk model
($r_d=4.0$kpc, $h_z=2.5$kpc, $\Sigma_0=90 \Msol/pc^2$) (solid line) and standard halo (dashed line).}
\end{figure}

\begin{figure}
\epsfysize=10.0cm
\centerline{
\rotate[r]{\epsfbox{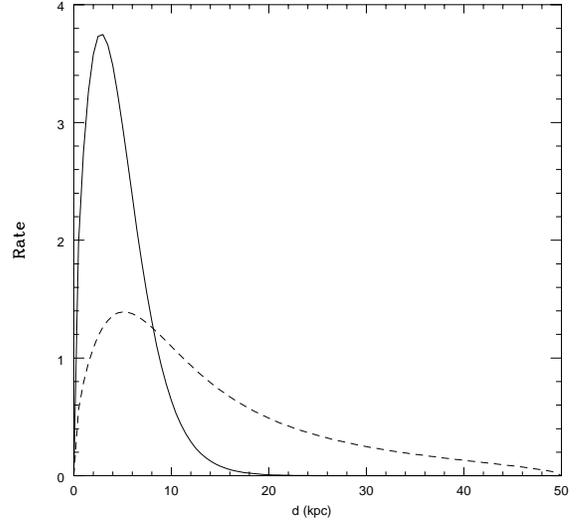}}}
\caption{Distribution of observer-lens distance for a typical thick
disk model $r_d=4.0$kpc, $h_z=2.5$kpc, $\Sigma_0=90 \Msol/pc^2$() (solid
line) and standard halo (dashed line).}
\end{figure}

\subsection{Parallax Events}

The canonical microlensing light curve is based on a number of
assumptions, among which is a constant transverse velocity. Since
microlensing events have a non-zero duration, the motion of the Earth in
its orbit around the Sun breaks this idealization. For most microlensing
events the resulting modification of the light curve is small. For longer
events, however, the effect of the Earth's motion is more apparent. The
detection of such a modification to the standard light curve is
interesting because it allows one to partially break the degeneracy
between lens distance, mass and velocity \cite{gould92}.

We follow Gould (1998) and define the quantity 
\be 
\gamma = \frac{v_\oplus}{\tilde{v}} \frac{2 \pi t_e}{{\rm yr}} \cos \phi,
\ee 
which measures the ratio of the earth's acceleration along $\tilde{v}$
(the projected lens velocity vector) during the event and
$\abs{\tilde{v}}$ itself. This gives us an approximate measure of the
strength of the parallax effect of an event. For a given event duration
the effe ct increases as the observer-lens distance and the typical lens
velocity decrease. Thus, a typical halo ($<v>\approx 200 {\rm km/s},
<x>=1/4$) event will have $\gamma\approx 0.05$ whereas if the MACHOs are
arranged in a thick disk configuration ($<v>\approx 100 {\rm km/s},
<x>=1/10$) then typically $\gamma \approx 0.11$.  We thus expect an
increase in the number of observed parallax events for lenses in a fat
disk configuration.

\begin{figure}
\epsfysize=10.0cm
\centerline{
\rotate[r]{\epsfbox{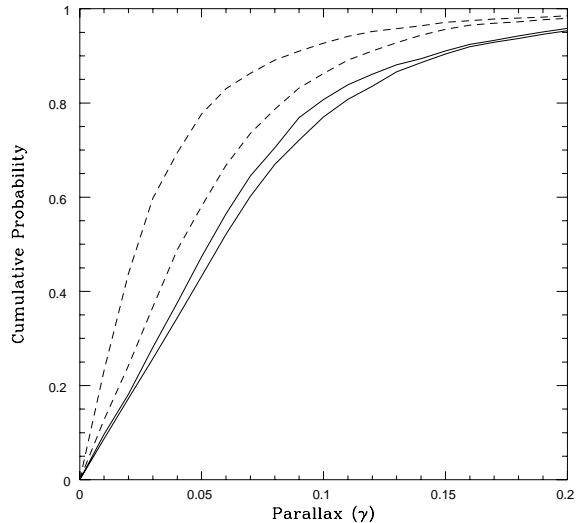}}}
\caption{Cumulative distribution of the $\gamma$ parallax parameter
for the halo (dashed lines) and thick disk (solid lines) scenarios.
Curves for both 1\% photometry (left) and 10\% photometry (right) are
shown. Neither blending nor backgrounds such as binary sources/lenses have
been taken into account.}
\end{figure}

In order to more carefully estimate the increase in expected parallax
events, we have performed a Monte Carlo analysis of lensing events for
lenses in a fat disk and a halo distribution. We sample the light curve
for each event in a manner chosen to correspond roughly to the present
surveys: daily measurements, 10\% photometry, and 5 year baseline, in order to
generate the data.  We also consider an experiment with 1\% photometry for
comparison.  We then fit a first order parallax light-curve to the data
and determine Gould's $\gamma$ parameter \cite{newgould} as a measure of
the strength of the parallax effects.

The distribution of measured $\gamma$'s can then be used to distinguish
between very thick disk and halo lens distributions.  In Figure 7 we
present the cumulative probability distributions for $\gamma$ for a thick
disk and a halo. The ability of an experiment to distinguish between these
lens distributions will depend on the number of events and photometry. We
find that an experiment with 1\% photometry and 15 events can distinguish
between a very thick disk and a halo with a significance of about 5\%,
while an experiment with 10\% photometry is unable to do so.  With 10\%
photometry at least 75 events are required.  We note that these estimates
do not include backgrounds from binary source and lens events, whose light
curves can mimic parallax effects. However, future observations are likely
to include more finely sampled light curves from followup data on alerted
events.  Such detailed light curves will increase the ability to
discriminate between parallax and binary lens or
source effects in at least some of the cases. 

\section{Conclusions}

Microlensing studies have yielded much exciting data in the past few years
and are continuing to survey different lines of sight through the Galaxy
in order to probe the Galactic halo.  However, the conclusions that can be
drawn from the data to date are very model dependent -- assumptions about
the distribution of the lenses and their velocity structure have a strong
impact on their interpretation. Thus we need to examine a wide range of
reasonable lens distributions.

Very thick disks present a reasonable alternative to a halo population of
lenses.  If the lenses are stellar remnants, it seems likely that their
configuration will be more condensed than that of a standard non-baryonic
halo.  While we have found that very thick disks cannot lower the lens
mass estimate to the brown dwarf regime, they have the advantage that
their total mass in MACHOs is somewhat less than that for a standard halo
that is truncated at $50\kpc$ (and much less than a MACHO halo which
traces the extended dark halo out to at least $100\kpc$.)  A thick disk
distribution cannot produce an optical depth toward the LMC of more than
about $1.5 \times 10^{-7}$.  However, it can explain (within the
experimental uncertainties) all or a significant fraction of the current
optical depth estimates.

As more events are detected, it may be possible to distinguish between
very thick disk and halo lens distributions.  The most promising avenue
for such a discriminant is the observation of parallax events.  Because
disk lenses would be both closer and on average slower than halo lenses we
expect a higher rate of such events for a disk population. Although the
survey experiment light curve measurements can only marginally
discriminate between disk and halo distributions for reasonable numbers of
events, a modest expenditure of telescope time to obtain one percent
photometry on all the Magellanic cloud events should be capable of making
the distinction very clearly.

\end{document}